\documentclass[journal]{IEEEtran}
\IEEEoverridecommandlockouts

\usepackage{cite}
\usepackage{graphicx}
\usepackage{amsmath,amssymb,amsfonts}
\usepackage{algorithmic}
\usepackage{textcomp}
\usepackage{xcolor}
\usepackage{lineno}

\begin{document}

\title{Blind Gradient-Ascent Phase Alignment for Multi-Aperture Coherent Digital Combining Under Aperture-Dependent Phase Disturbance}

\author{Cheng~Chen,~Tong~Luo,~Jiayin~Xue,~Siyu~Gong,~Qun~Zhang,~
        Linsheng~Fan,~Qi~Wu,~and~Yanfu~Yang
\thanks{Manuscript received xxxx xx, 2025; revised xxxx xx, 2025; accepted xxxx xx, 2025. Date of publication xxxx xx, 2025; date of current version xxxx xx, 2025.}
\thanks{C. Chen, T. Luo, J. Xue, S. Gong, Q. Zhang, L. Fan, and Q. Wu are with Pengcheng Laboratory, Shenzhen 518000, China. (e-mail: chengch01@pcl.ac.cn)}
\thanks{Y. Yang is with the Department of Electronic and Information Engineering, Harbin Institute of Technology, Shenzhen 518055, China. (email: yangyanfu@hit.edu.cn).}
}

\markboth{Journal of Lightwave Technology,~Vol.~XX, No.~XX, XXXX~2025}%
{Chen \MakeLowercase{\textit{et al.}}: Blind Gradient-Ascent Phase Alignment for Multi-Aperture Coherent Digital Combining}

\maketitle

\begin{abstract}
Multi-aperture reception can provide spatial diversity in free-space optical (FSO) communication by collecting signal replicas at separate apertures. If the received fields are also phase-aligned, the same architecture can provide coherent-combining gain. This paper proposes blind gradient-ascent phase alignment (BGAPA), which updates one scalar phase correction per aperture by maximizing the combined output power. The update uses a closed-form analytical gradient and does not require symbol decisions, unlike SPGD with stochastic perturbation-based gradient estimation or DD-LMS with decision-directed feedback. To focus on phase tracking, the numerical model includes independent aperture-dependent phase disturbance but excludes amplitude scintillation and polarization-dependent distortion. In this controlled phase-only setting, increasing the aperture count from 64 to 256 gives BGAPA an SNR improvement of about 5.7~dB, close to the ideal 6.02~dB gain for a fourfold increase in aperture count. At $N=16$ and $f_{\max}=1$~MHz, BGAPA maintains an almost flat post-CPR SNR across randomly sampled $A_{\text{phase}} \in [2,500]$~rad. When $f_{\max}$ is randomly sampled across 1.4--98~MHz, BGAPA also gives higher mean SNR than the benchmark schemes at both $N=16$ and $N=64$. With 2-sps fourth-power frequency-offset estimation, the receiver remains below the HD-FEC threshold for common carrier-frequency offsets up to $\lvert\Delta f\rvert\leq6$~GHz. The reported step size is optimized separately at each operating point. BGAPA updates its phase parameters directly from the received aperture fields, without training symbols, pilots, or decision-directed feedback.
\end{abstract}


\IEEEpeerreviewmaketitle

\section{Introduction}
\label{sec:intro}

\IEEEPARstart{F}{ree-space} optical (FSO) communication offers high carrier frequency, narrow beam divergence, immunity to electromagnetic interference, and high spatial reuse, making it attractive for terrestrial access links, inter-building links, and satellite-to-ground downlinks~\cite{Khalighi2014,Kaushal2017}. Coherent detection further improves receiver sensitivity and enables advanced digital signal processing (DSP), which is important for power-constrained long-distance links~\cite{Ip2008}. Atmospheric turbulence can distort the received optical field through intensity scintillation, phase distortion, and polarization rotation~\cite{Zhu2002}. This work focuses on the aperture-dependent phase component of this problem.

Several receiver-side techniques have been developed to mitigate turbulence-induced fading. Aperture averaging reduces scintillation by increasing the collection area~\cite{Andrews2001}, but a larger aperture also captures stronger spatial phase variation across the receiver pupil, making efficient single-mode coupling difficult without adaptive optics (AO)~\cite{Tyson2015}. AO can correct wavefront distortion, yet its cost, spatial resolution, and loop bandwidth limit its practicality under strong or rapidly varying turbulence. Multi-aperture reception provides an alternative path based on spatial diversity: multiple smaller apertures are deployed at the same receiving terminal and arranged so that the turbulence-induced fading observed by different apertures is weakly correlated or approximately independent~\cite{Lee2002}. This architecture reduces the probability that all branches experience deep fading simultaneously, improving link stability, and it is naturally compatible with digital coherent receivers provided that the aperture branches can also be combined coherently.

The key difficulty in multi-aperture coherent digital combining is not only collecting multiple replicas, but also aligning their complex phases before summation. Spatial diversity improves robustness by exploiting weakly correlated fading across apertures. Coherent-combining gain, however, is obtained only when the optical fields are co-phased and added constructively. In practice, different branches can exhibit static skew, unequal received power, and time-varying carrier phase offsets~\cite{Liu2023}. A recent real-time two-aperture experiment showed that strong turbulence can induce high-dynamic phase fluctuations that reduce combining efficiency if left uncompensated~\cite{Ju2024}. Conventional diversity rules such as maximal-ratio combining (MRC), equal-gain combining (EGC), and selection combining (SC) have been used in FSO links~\cite{Navidpour2007}. Coherent MRC uses channel-dependent complex weights that include phase compensation. Under the equal-per-aperture-SNR condition studied here, the magnitude weights reduce to equal scaling, while phase alignment remains necessary before constructive summation. The present work compares phase-tracking mechanisms under identical per-aperture signal and noise conditions.

Digital phase-alignment methods have become central to multi-aperture coherent combining. Geisler~\textit{et al.} demonstrated a receiver in which relative phase offsets are estimated by block-wise cross-correlation and compensated digitally, achieving coherent combining of four apertures at very low received power~\cite{Geisler2016}. The block-wise estimate benefits from coherent averaging, but the block length also limits its response to rapid phase changes. Stochastic parallel gradient descent (SPGD)~\cite{Chang2020} offers a blind alternative by perturbing all phase controls and estimating the power gradient from the resulting metric change. The resulting gradient estimate contains perturbation noise, which can slow adaptation when the aperture count is large or the phase disturbance varies rapidly. MIMO equalizer-based approaches, including complex-valued $2N \times 2$ CMA equalization~\cite{Liu2023}, real-valued $4N \times 2$ FPGA implementation~\cite{Ju2024}, and frequency-domain overlap-save equalization~\cite{Chen2025}, adapt many FIR coefficients to recover the combined signal. CMA and RDE operate through blind modulus- or radius-based error criteria, which are indirect objectives for coherent combining. Decision-directed LMS (DD-LMS) combining has also been used for coherent multi-aperture reception, where the complex combining weights are adapted from the error between the combined signal and the detected symbol~\cite{Xu2023}. Unlike CMA/RDE radius decisions, DD-LMS relies on symbol decisions, so unreliable decisions can affect the update when residual phase error is large. These equalizer-based methods are flexible, but phase-dominant aperture mismatch can often be represented by one phase variable per aperture.

This paper develops a blind phase combiner for the phase-dominant setting. The proposed blind gradient-ascent phase alignment (BGAPA) models each aperture by a scalar phase correction and updates these corrections by maximizing the instantaneous combined output power. The update uses a closed-form analytical gradient computed from the received aperture fields and the combined output. This avoids both random perturbations and symbol-decision feedback during phase-parameter adaptation.

The remainder of this paper is structured as follows. Section~II describes the principle of the proposed method, including the signal parameterization and definition of the objective function. Section~III derives the closed-form gradient expressions and presents the algorithm pseudocode. Section~IV details the numerical simulation setup and evaluates combining performance, OSNR penalty, disturbance-frequency robustness, disturbance-amplitude tolerance, and tolerance to a common carrier-frequency offset. Section~V concludes the paper.

\section{Principle of the Proposed Method}
\label{sec:principle}

We consider a multi-aperture coherent free-space optical receiver comprising $N$ independent apertures, each followed by a coherent detector that recovers both orthogonal polarization components of the received optical field. For a dual-polarization transmission format, the complex baseband signal vector captured by the $n$-th aperture at a given sampling instant is denoted by
\begin{equation}
\mathbf{E}_n =
\begin{bmatrix}
E_{x,n} \\
E_{y,n}
\end{bmatrix},
\label{eq:En}
\end{equation}
where $E_{x,n}, E_{y,n} \in \mathbb{C}$ represent the complex-valued baseband samples in the $x$ and $y$ polarization branches, respectively. Atmospheric turbulence introduces time-varying phase offsets across the $N$ apertures. To compensate for these impairments prior to coherent combining, we apply a per-aperture scalar phase correction. Let $\phi_n \in \mathbb{R}$ denote the phase correction applied to the $n$-th aperture. The phase-corrected signal vector is
\begin{equation}
\mathbf{U}_n = e^{j\phi_n} \mathbf{E}_n =
\begin{bmatrix}
U_{x,n} \\
U_{y,n}
\end{bmatrix}
=
\begin{bmatrix}
E_{x,n} e^{j\phi_n} \\
E_{y,n} e^{j\phi_n}
\end{bmatrix}.
\label{eq:Un}
\end{equation}

The coherently combined signals in the two orthogonal polarizations are obtained by summation across all $N$ apertures:
\begin{align}
S_x &= \sum_{n=1}^{N} U_{x,n}, \label{eq:Sx} \\
S_y &= \sum_{n=1}^{N} U_{y,n}. \label{eq:Sy}
\end{align}

To guide the adaptation of the phase shifts $\{\phi_n\}_{n=1}^N$, we adopt the total instantaneous output power as the maximization objective:
\begin{equation}
J(\phi_1,\dots,\phi_N) = |S_x|^2 + |S_y|^2.
\label{eq:J}
\end{equation}

The objective in (\ref{eq:J}) follows the coherent-combining criterion: when the aperture fields are phase-aligned, the summed field power increases. The same objective is differentiable with respect to $\{\phi_n\}_{n=1}^N$, which allows gradient-based phase adaptation without training symbols, pilot sequences, or demodulator feedback.

\section{Blind Gradient-Ascent Phase Optimization}
\label{sec:gradient}

\subsection{Gradient Derivation}

We derive the partial derivative of $J$ with respect to the phase $\phi_n$. Since $J = |S_x|^2 + |S_y|^2$, the gradient decomposes as
\begin{equation}
\frac{\partial J}{\partial \phi_n}
= \frac{\partial |S_x|^2}{\partial \phi_n} + \frac{\partial |S_y|^2}{\partial \phi_n}.
\label{eq:dJ_dphi_decompose}
\end{equation}
Differentiating $|S_x|^2 = S_x S_x^*$ and applying the product rule yields
\begin{equation}
\frac{\partial |S_x|^2}{\partial \phi_n}
= 2\,\operatorname{Re}\!\left(\frac{\partial S_x}{\partial \phi_n} S_x^*\right).
\label{eq:dSx2_compact}
\end{equation}

Applying the same derivation to $|S_y|^2$ and summing both contributions, the total gradient is
\begin{equation}
\frac{\partial J}{\partial \phi_n}
= 2\,\operatorname{Re}\!\left(
\frac{\partial S_x}{\partial \phi_n} S_x^* +
\frac{\partial S_y}{\partial \phi_n} S_y^*
\right).
\label{eq:dJ_dphi_general}
\end{equation}

From (\ref{eq:Sx})--(\ref{eq:Sy}), only the $n$-th aperture terms depend on $\phi_n$; therefore
\begin{equation}
	\begin{split}
\frac{\partial S_x}{\partial \phi_n} = \frac{\partial U_{x,n}}{\partial \phi_n} = j U_{x,n},\\
\frac{\partial S_y}{\partial \phi_n} = \frac{\partial U_{y,n}}{\partial \phi_n} = j U_{y,n}.
\label{eq:dSn_dphi}
\end{split}
\end{equation}

Substituting (\ref{eq:dSn_dphi}) into (\ref{eq:dJ_dphi_general}) and simplifying, we obtain the closed-form phase gradient
\begin{equation}
\frac{\partial J}{\partial \phi_n}
= -2\,\operatorname{Im}\!\left(
U_{x,n} S_x^* + U_{y,n} S_y^*
\right).
\label{eq:dJ_dphi_final}
\end{equation}

The corresponding gradient-ascent update rule for the phase parameter is
\begin{equation}
\begin{aligned}
\phi_n^{(k+1)} &= \phi_n^{(k)} + \gamma \frac{\partial J}{\partial \phi_n} \\
&= \phi_n^{(k)} - 2\gamma \,\operatorname{Im}\!\left(
U_{x,n} S_x^* + U_{y,n} S_y^*
\right),
\end{aligned}
\label{eq:phi_update}
\end{equation}
where $\gamma > 0$ is the step size and $k$ denotes the iteration index.

\subsection{Algorithm Summary}

\begin{algorithmic}[1]
\STATE {\bf Input:} Multi-aperture signal $\{\mathbf{E}_n(k)\}_{n=1}^N$, step size $\gamma$
\STATE {\bf Output:} Combined signal $S_x(k)$, $S_y(k)$
\STATE Initialize $\phi_n \leftarrow 0$ for $n = 1,\dots,N$
\FOR{each sample $k = 1, 2, \dots$}
    \STATE $S_x \leftarrow 0$, \quad $S_y \leftarrow 0$
    \FOR{each aperture $n = 1,\dots,N$ (in parallel)}
        \STATE $\mathbf{U}_n \leftarrow e^{j\phi_n}\,\mathbf{E}_n(k)$ \COMMENT{Phase correction}
        \STATE $S_x \leftarrow S_x + U_{x,n}$, \quad $S_y \leftarrow S_y + U_{y,n}$
    \ENDFOR
    \FOR{each aperture $n = 1,\dots,N$ (in parallel)}
        \STATE $g_n \leftarrow -2\,\operatorname{Im}\!\bigl(U_{x,n}S_x^* + U_{y,n}S_y^*\bigr)$ \COMMENT{Gradient}
        \STATE $\phi_n \leftarrow \phi_n + \gamma \cdot g_n$ \COMMENT{Update}
    \ENDFOR
\ENDFOR
\end{algorithmic}

The procedure uses no training symbols or pilot sequences and operates on a per-sample basis. In the simulation implementation, the current samples are used to evaluate the gradient, and the updated phases are then applied to the buffered current samples to form the reported combined output. The gradient in (\ref{eq:dJ_dphi_final}) is exact for the instantaneous objective in (\ref{eq:J}), but this local property does not prove global convergence under time-varying phase disturbance and additive noise. The step size is selected empirically by the operating-point-specific sweeps described below.

\section{Numerical Simulation and Results}
\label{sec:simulation}

\subsection{Simulation Setup}

\begin{figure}[!htbp]
\centering
\includegraphics[width=\linewidth]{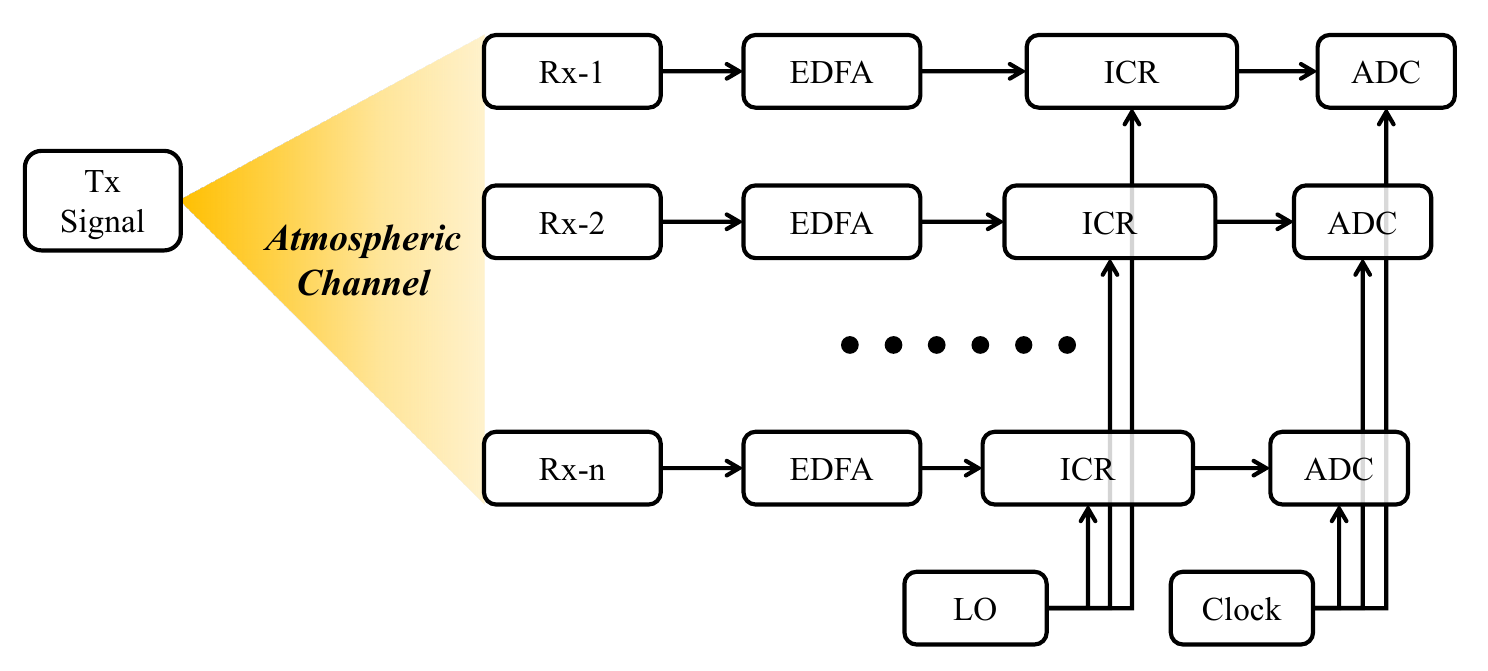}
\caption{Simulation framework used for the phase-only multi-aperture benchmark.}
\label{fig:framework}
\end{figure}

We evaluate BGAPA and four benchmark schemes through numerical simulation of a coherent multi-aperture receiver subject to synthetic aperture-dependent phase disturbance. The simulation includes the transmitter, phase-disturbance model, receiver DSP chain, and performance evaluation stages illustrated in Fig.~\ref{fig:framework}.

The channel model deliberately isolates synthetic aperture-dependent phase disturbance. Scintillation-induced amplitude fading, polarization-dependent distortion, and spatial or temporal statistics derived from a physical turbulence model are not included. The resulting controlled phase-only setting is used to compare the per-aperture phase-tracking capability of the combining algorithms and should not be interpreted as a complete atmospheric-channel simulation.

The transmitter generates dual-polarization QPSK symbols at a symbol rate of $R_s = 25$~Gbaud. Each polarization branch independently produces $L = 2^{15}$ random QPSK symbols drawn from the constellation set $\{\pm 1 \pm j\}$. Transmit symbols are pulse-shaped with a root-raised-cosine (RRC) filter with a roll-off factor $\beta = 0.2$ and a filter span of 128 symbols, operating at $N_{\text{sps}} = 2$ samples per symbol. The transmit and local-oscillator lasers are each modeled with a Lorentzian linewidth of 100~kHz; their aggregate carrier phase noise is superimposed on the aperture-dependent phase disturbance.

To emulate time-varying atmospheric phase distortion on each aperture, we use a superposition of multiple sinusoidal tones. For the $n$-th aperture, the time-varying phase is generated as
\begin{equation}
\varphi_n(t) = \sum_{j=1}^{N_{\text{tones}}} A_{n,j} \sin\!\bigl(2\pi f_j t + \psi_{n,j}\bigr),
\label{eq:turb_phase}
\end{equation}
where the $j$-th tone frequency is $f_j = j \cdot f_{\text{max}} / N_{\text{tones}}$, the amplitude takes the form $A_{n,j} = A_0 \cdot s_{n,j}$ with $A_0 = A_{\text{phase}} \sqrt{2 / N_{\text{tones}}}$ and a random binary sign $s_{n,j} \in \{\pm 1\}$, the initial phase $\psi_{n,j} \sim \mathcal{U}(0, 2\pi)$, and all random variables are independently drawn per aperture. In the amplitude-tolerance experiment, the phase RMS is calculated separately for each aperture over time and then averaged across all apertures.

\begin{figure}[!htbp]
\centering
\includegraphics[width=\linewidth]{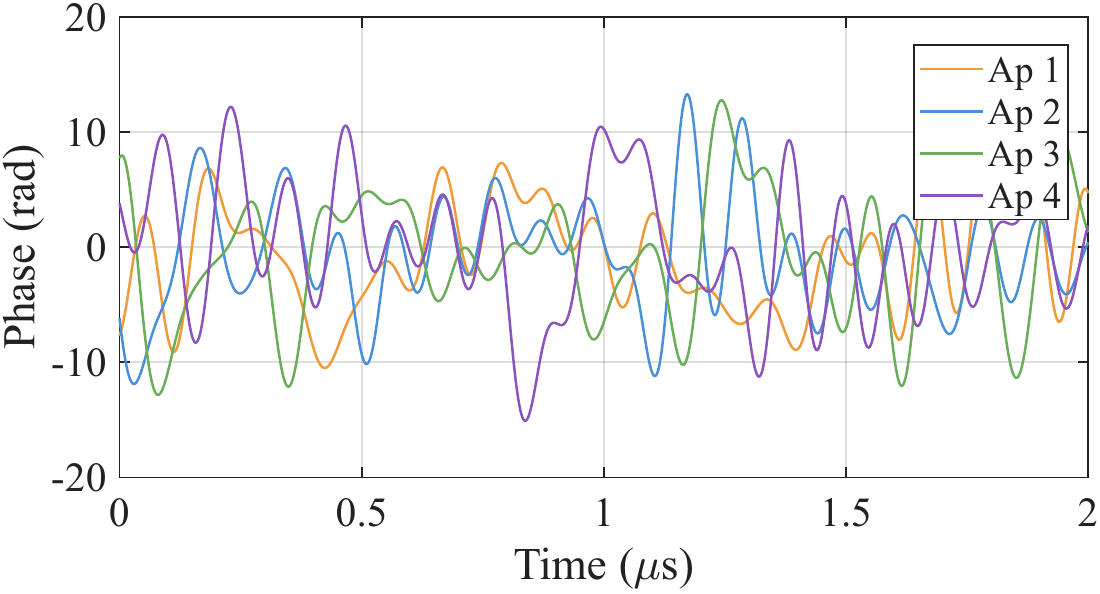}
\caption{Representative multi-tone phase disturbance waveforms of four apertures over a 2-$\mu$s window at $A_{\text{phase}} = 5$~rad, $f_{\text{max}} = 10$~MHz, and $N_{\text{tones}} = 64$.}
\label{fig:phase_noise}
\end{figure}

Fig.~\ref{fig:phase_noise} shows representative phase disturbance waveforms generated by (\ref{eq:turb_phase}) for four independently generated apertures at $A_{\text{phase}} = 5$~rad and $f_{\text{max}} = 10$~MHz. Their distinct trajectories result from the independent random signs $s_{n,j}$ and initial phases $\psi_{n,j}$.

At the receiver, the optical signal collected by each aperture is amplified prior to coherent detection, introducing amplified spontaneous emission (ASE) noise. Unless otherwise specified, the baseline OSNR is set to $\mathrm{OSNR} = 10$~dB referenced to a 12.5~GHz noise bandwidth, corresponding to the 0.1~nm wavelength resolution at the C-band operating wavelength of 1550~nm.

After coherent detection, the electrical signals are passed through a matched RRC filter. These multi-aperture matched-filtered signals are then processed and coherently combined by the proposed method and four benchmark schemes:
\begin{enumerate}
\item \textbf{Block-wise cross-correlation}~\cite{Geisler2016}: relative phases estimated via complex dot products over blocks of $M$ samples with a sequentially accumulated coherent reference:
\begin{equation}
\varphi_i = \arg\!\left(\sum_{m=1}^{M} S_i[m] \cdot \overline{U_{i-1}[m]}\right),
\label{eq:geisler}
\end{equation}
where the overbar denotes complex conjugation, and the running coherent sum is updated as $U_1 = S_1$ and $U_i = U_{i-1} + e^{-j\varphi_i} S_i$ for $i = 2, \dots, N$, yielding the final combined output $U_N$.
\item \textbf{SPGD}~\cite{Chang2020}: stochastic parallel gradient descent with binary perturbations:
\begin{equation}
\begin{gathered}
\boldsymbol{\xi} = \operatorname{sign}(\mathbf{u}), \quad \mathbf{u} \sim \mathcal{U}(-1, 1)^N, \\[2pt]
J^\pm =
	\Bigl\|\sum_{n=1}^{N} \mathbf{E}_n \, e^{j(\boldsymbol{\phi} \pm \delta \boldsymbol{\xi})}\Bigr\|^2, \\[2pt]
\boldsymbol{\phi}^{(k+1)} = \boldsymbol{\phi}^{(k)} + \gamma \left(J^+ - J^-\right) \boldsymbol{\xi},
\end{gathered}
\label{eq:spgd}
\end{equation}
where $\boldsymbol{\xi} = [\xi_1,\dots,\xi_N]^T$ with $\xi_n \in \{\pm 1\}$ is a random binary perturbation vector and $\delta = \pi/12$ is the fixed perturbation magnitude.
\item \textbf{CMA/RDE}~\cite{Liu2023,Ju2024}: single-tap per-aperture CMA/RDE combining, without polarization demultiplexing. For QPSK, the RDE radius decision reduces to the constant-modulus reference, so CMA and RDE are represented by a single benchmark curve. The cost function and stochastic gradient update are~\cite{Mello2021}:
\begin{equation}
\begin{gathered}
J_{\text{CMA}} = \mathbb{E}\!\left[\left(|S|^2 - R_d\right)^2\right], \\[2pt]
w_n^{(k+1)} = w_n^{(k)} + \mu \left(R_d - |S_k|^2\right) S_k^* \, E_n(k),
\end{gathered}
\label{eq:cma}
\end{equation}
where $S_k = \sum_{n=1}^{N} w_n^{(k)} E_n(k)$ and $R_d = \mathbb{E}[|a|^4] / \mathbb{E}[|a|^2]$.
\item \textbf{DD-LMS}~\cite{Xu2023}: decision-directed LMS combining that jointly performs phase alignment and carrier phase recovery via $N$ complex weights:
\begin{equation}
\begin{gathered}
J_{\text{DD-LMS}} = \mathbb{E}\!\left[|\hat{a}_k - S_k|^2\right], \\[2pt]
w_n^{(k+1)} = w_n^{(k)} + \mu \left(\hat{a}_k - S_k\right) E_n^*(k),
\end{gathered}
\label{eq:ddlms}
\end{equation}
where $S_k = \sum_{n=1}^{N} w_n^{(k)} E_n(k)$ and $\hat{a}_k = \operatorname{dec}(S_k)$ is the symbol decision.
\end{enumerate}

All methods are evaluated in the same phase-only benchmark setting: each aperture has the same signal and noise statistics, and each branch applies a single scalar weight $w_n \in \mathbb{C}$ (or a pure phase rotation $e^{j\phi_n}$) to both polarization components. Since the simulation does not include polarization-dependent impairments, the X and Y branches experience the same phase offset at each aperture and share the same combining weight. The CMA/RDE and DD-LMS implementations are therefore single-tap phase-tracking benchmarks rather than reproductions of complete multi-tap polarization-demultiplexing architectures. The key parameter of every scheme, including the BGAPA step size $\gamma$, is optimized separately by a parameter sweep at each reported operating point.

The update mechanisms differ despite this common benchmark scope. SPGD~\cite{Chang2020} estimates the gradient through random perturbations and scalar power changes, so its direction contains perturbation noise. DD-LMS~\cite{Xu2023} uses a decision-directed error signal whose reliability can decrease when residual phase error causes decision errors. CMA/RDE-based adaptation~\cite{Liu2023,Ju2024,Chen2025} minimizes a modulus-error criterion rather than the instantaneous combined power. BGAPA instead uses the analytical gradient of the instantaneous power objective. The comparison below is limited to this phase-only benchmark and does not claim global convergence or general superiority outside the tested conditions.

The aperture-combining algorithms primarily compensate the relative phase disturbances among the received branches. In contrast, the laser phase noise is common to all apertures and, together with any common residual phase error left after aperture alignment, must be tracked at the combined output. For BGAPA, SPGD, and CMA/RDE, a sliding-window Viterbi--Viterbi carrier phase recovery (CPR) stage is applied after combining to estimate and compensate this common phase term. The block-wise cross-correlation scheme is followed by a block-wise Viterbi--Viterbi CPR stage whose processing interval is matched to the combining block. DD-LMS is treated differently in the main aperture-count and OSNR sweeps: its complex combining weights are updated from the detected-symbol error and therefore jointly track the inter-aperture phase offsets and the common carrier phase; no additional CPR stage is applied to its output. Thus, the reported SNR and BER are evaluated after the corresponding phase-recovery operation for each scheme and subsequent phase-ambiguity resolution. Unless otherwise stated, the results are averaged over 50 independent Monte Carlo trials.

\subsection{Combining Performance}

\begin{figure}[!htbp]
\centering
\includegraphics[width=\linewidth]{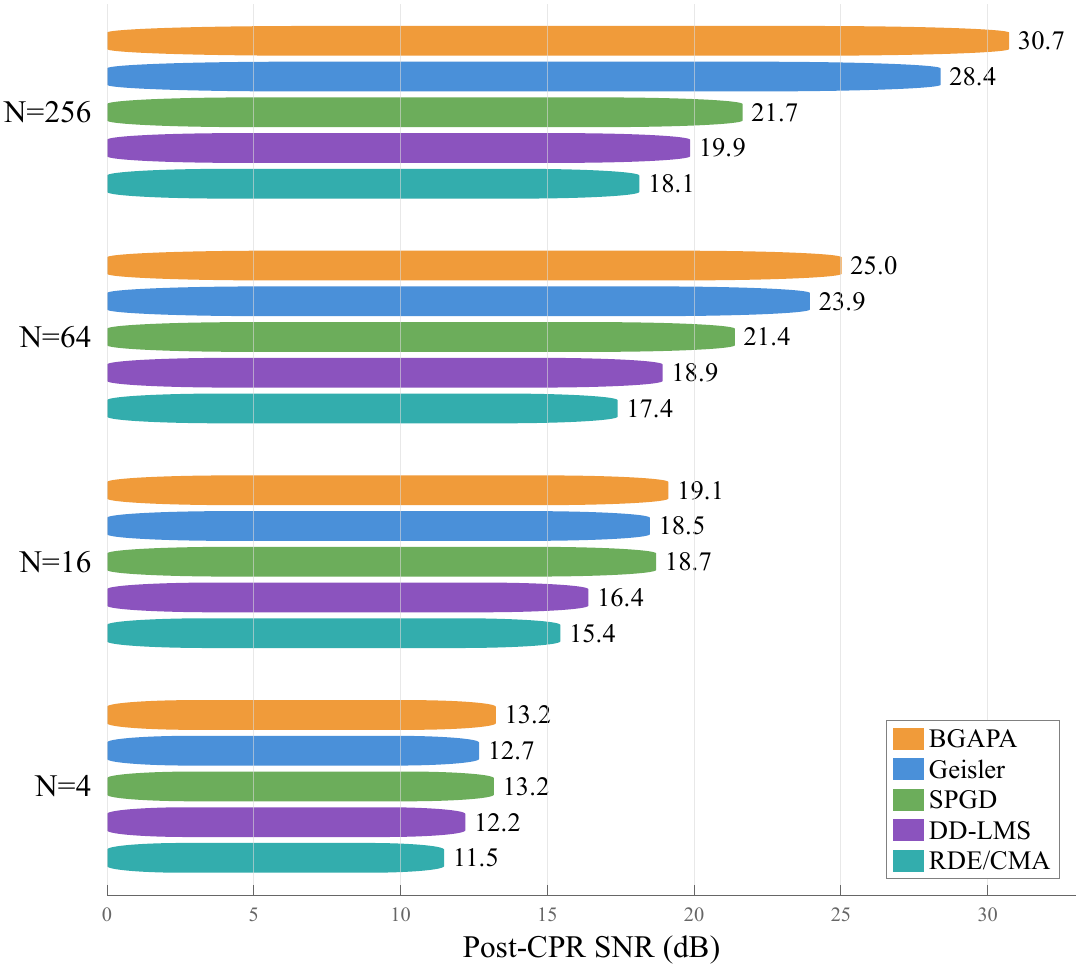}
\caption{Post-CPR SNR versus aperture count under phase disturbance. Results are Monte Carlo means over 50 trials with $f_{\text{max}} = 10$~MHz, $A_{\text{phase}} = 5$~rad, and $\mathrm{OSNR} = 10$~dB.}
\label{fig:snr_vs_n}
\end{figure}

We first examine how the combining gain scales with the number of receiver apertures. Fig.~\ref{fig:snr_vs_n} shows the post-CPR SNR of the five plotted schemes, including the proposed method, as a function of the aperture count $N_{\text{ap}}$ at a fixed $f_{\text{max}} = 10$~MHz, $A_{\text{phase}} = 5$~rad, and $\mathrm{OSNR} = 10$~dB.

All methods show higher SNR as the number of apertures increases, but their scaling differs. BGAPA gives the highest SNR at each tested aperture count. Its gap relative to SPGD and Geisler's method is small at $N_{\text{ap}} = 4$ and becomes larger at higher aperture counts. DD-LMS remains above the CMA/RDE benchmark, while CMA/RDE falls further behind as $N_{\text{ap}}$ increases. When $N_{\text{ap}}$ is increased from 64 to 256, BGAPA gains about 5.7~dB, close to the ideal fourfold coherent-combining gain of $10\log_{10}(4)=6.02$~dB.

\subsection{OSNR Penalty Analysis}

\begin{figure}[!htbp]
\centering
\includegraphics[width=\linewidth]{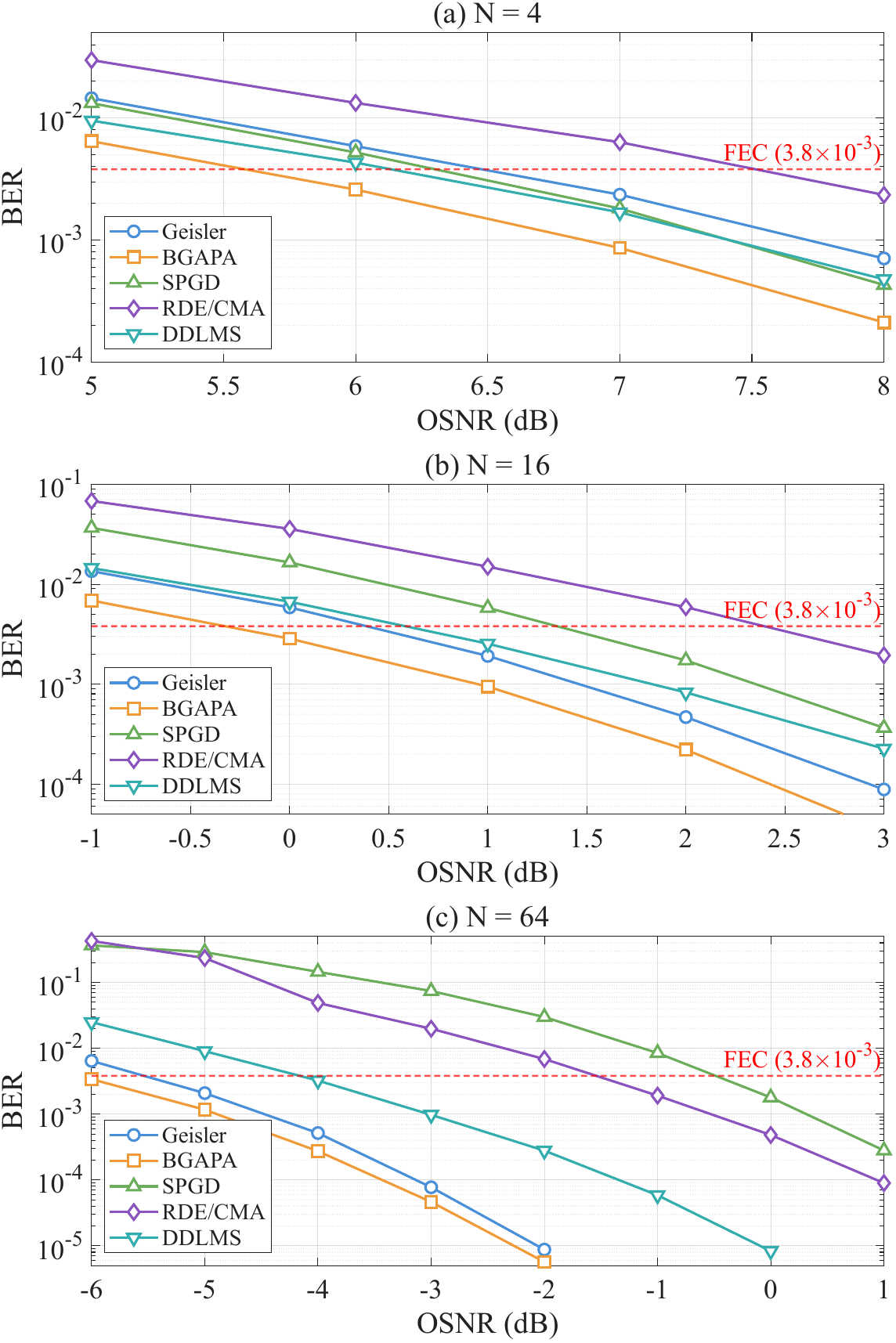}
\caption{BER versus OSNR under phase disturbance with $f_{\text{max}} = 10$~MHz and $A_{\text{phase}} = 5$~rad. For QPSK, CMA and RDE are represented by one RDE/CMA benchmark curve. (a) $N_{\text{ap}} = 4$. (b) $N_{\text{ap}} = 16$. (c) $N_{\text{ap}} = 64$. The dashed red line marks the HD-FEC threshold at $\mathrm{BER} = 3.8 \times 10^{-3}$. Each BER point is a trimmed mean over 10 trials after discarding one worst-BER trial.}
\label{fig:ber_vs_osnr}
\end{figure}

To connect the SNR gain with receiver sensitivity, we sweep the OSNR and measure the BER. Fig.~\ref{fig:ber_vs_osnr} shows the BER versus OSNR curves of the proposed method and four benchmark schemes for aperture counts of $N_{\text{ap}} = 4$, $16$, and $64$ under phase disturbance with $f_{\text{max}} = 10$~MHz and $A_{\text{phase}} = 5$~rad. The OSNR is referenced to a 12.5~GHz noise bandwidth. At each OSNR point, BER is computed by direct error counting over $L = 2^{16}$ QPSK symbols in 10 independent Monte Carlo trials; the plotted value is a trimmed mean obtained after discarding the single worst-BER trial for that method and OSNR.

At $N_{\text{ap}} = 4$, BGAPA reaches the hard-decision forward error correction (HD-FEC) threshold of $\mathrm{BER} = 3.8 \times 10^{-3}$ at approximately 5.7~dB. Geisler's method and SPGD reach the same threshold at slightly higher OSNR values, while DD-LMS and the RDE/CMA benchmark require a larger OSNR margin. The gap among the better-performing methods is still modest at this aperture count, as also seen in Fig.~\ref{fig:snr_vs_n}.

At larger aperture counts, the OSNR gap increases. At $N_{\text{ap}} = 16$, BGAPA reaches the FEC threshold at approximately $-0.3$~dB. Geisler's method, SPGD, and the RDE/CMA benchmark cross the threshold at higher OSNR values. At $N_{\text{ap}} = 64$, BGAPA remains below the FEC threshold down to the lowest measured OSNR of $-6$~dB, whereas the benchmark schemes are above the threshold at this point. This trend is consistent with the aperture-count result in Fig.~\ref{fig:snr_vs_n}: the deterministic gradient is not affected by perturbation noise and does not require block averaging.

The BER curves follow the same ranking as the post-CPR SNR results, linking the SNR gain in Fig.~\ref{fig:snr_vs_n} to receiver sensitivity in this phase-only benchmark.

\subsection{Disturbance Frequency Robustness}

The preceding results are obtained at a single fixed disturbance frequency ($f_{\text{max}} = 10$~MHz). To assess robustness across a broader range of disturbance speeds, we sample $f_{\text{max}}$ randomly on a logarithmic scale. For each of 500 independent trials, the disturbance frequency is drawn as $f_{\text{max}} = 10^{a}$ with $a \sim \mathcal{U}(6 + 2\sqrt{c}, 6 + 2)$ where $c \sim \mathcal{U}(0,1)$, producing a distribution with a higher density toward the upper end of the 1.4--98~MHz range to stress methods at the fastest disturbance speeds. Each trial executes a full per-method parameter scan identical to that described in Sections~IV-B and IV-C, and the best post-CPR SNR is recorded for each method at $N_{\text{ap}} = 16$ and $N_{\text{ap}} = 64$ with $\mathrm{OSNR} = 5$~dB.

\begin{figure}[!htbp]
\centering
\includegraphics[width=\linewidth]{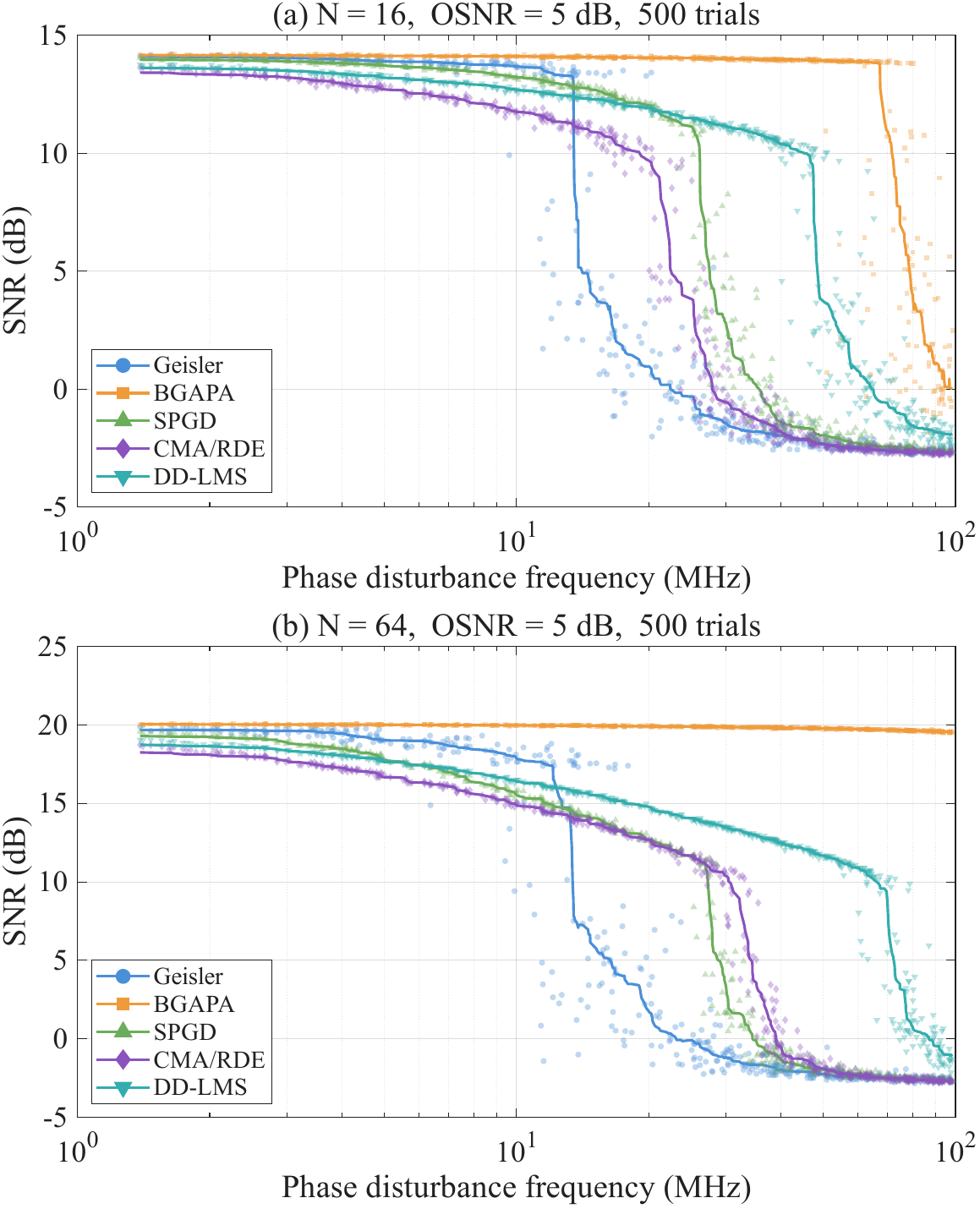}
\caption{Robustness to randomly sampled disturbance frequency $f_{\text{max}}$. Each point is the best post-CPR SNR of one trial after per-method parameter optimization. Moving-median trend lines (window: 12\% of trials) show the central trend. (a)~$N = 16$. (b)~$N = 64$. $\mathrm{OSNR} = 5$~dB and 500 trials per panel.}
\label{fig:random_fphase}
\end{figure}

Fig.~\ref{fig:random_fphase} presents the per-trial best SNR as a scatter plot overlaid with a moving-median trend line for each of the five evaluated schemes.

At $N_{\text{ap}} = 16$, BGAPA achieves a mean best SNR of 12.14~dB across 500 trials, exceeding DD-LMS by 3.7~dB, SPGD by 6.3~dB, and Geisler's method by 8.4~dB. The BGAPA standard deviation is 4.41~dB. Its moving-median trend remains above 10~dB over most of the tested range and declines at high $f_{\text{max}}$, where the disturbance approaches the tracking bandwidth of the update.

At $N_{\text{ap}} = 64$, the mean best SNR of BGAPA is 19.85~dB, exceeding DD-LMS by 7.4~dB, SPGD by 12.5~dB, and Geisler's method by 14.4~dB. The BGAPA standard deviation also drops to 0.14~dB, while the benchmark methods show standard deviations between 5.8 and 9.5~dB and degrade at high $f_{\text{max}}$.

The two aperture counts show different sensitivity to fast disturbance. At $N_{\text{ap}} = 16$, BGAPA still leads the benchmarks but loses margin at the highest disturbance speeds. At $N_{\text{ap}} = 64$, the trend is flatter across the sampled range. This behavior is consistent with the SNR scaling in Fig.~\ref{fig:snr_vs_n} and the OSNR ranking in Fig.~\ref{fig:ber_vs_osnr}.

\subsection{Phase Disturbance Amplitude Tolerance}

To evaluate tolerance to the disturbance amplitude, we sample $A_{\text{phase}}$ uniformly from 2 to 500~rad across 500 independent trials at $N_{\text{ap}} = 16$, $f_{\text{max}} = 1$~MHz, and $\mathrm{OSNR} = 5$~dB. For each trial, the step size (or equivalent control parameter) of every method is calibrated individually by a full parameter sweep, and the selected post-CPR SNR is recorded together with the actual per-aperture phase RMS measured from the generated disturbance waveform. The resulting curves represent performance after operating-point-specific parameter optimization.

\begin{figure}[!htbp]
\centering
\includegraphics[width=\linewidth]{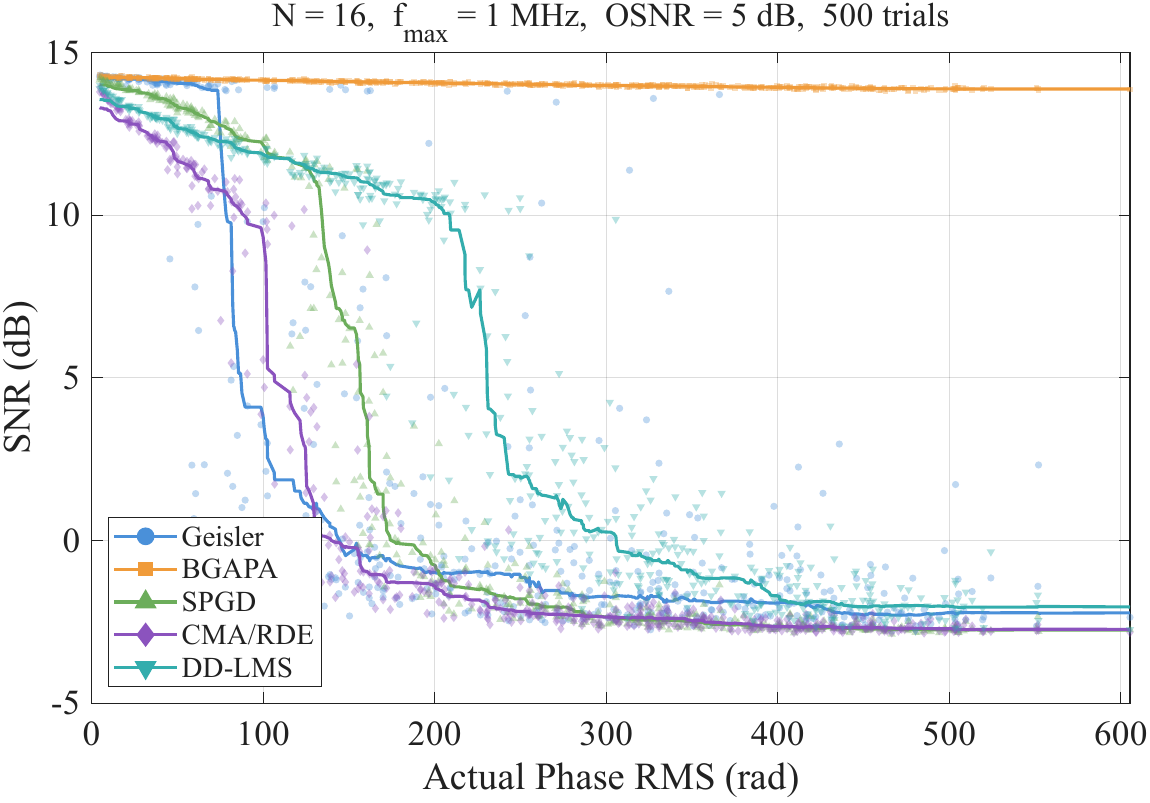}
\caption{Phase-disturbance amplitude tolerance under randomly sampled $A_{\text{phase}}$. Each point is the best post-CPR SNR of one trial after per-method parameter optimization. The horizontal axis gives the measured per-aperture phase RMS. $N_{\text{ap}} = 16$, $f_{\text{max}} = 1$~MHz, $\mathrm{OSNR} = 5$~dB, 500 trials.}
\label{fig:aphase_snr}
\end{figure}

Fig.~\ref{fig:aphase_snr} presents the per-trial best SNR as a scatter plot overlaid with a moving-median trend line for all five evaluated schemes.

BGAPA gives a nearly flat SNR trend across the tested amplitude range. Its mean best SNR over 500 trials is 14.04~dB, with a standard deviation of 0.11~dB. The moving-median trend remains almost unchanged from approximately 5~rad to more than 600~rad of measured phase RMS, provided that the step size is selected for the operating point.

The benchmark methods lose SNR as the disturbance amplitude increases. Around 300~rad RMS, their binned mean SNR values are approximately $-2.1$~dB (CMA/RDE), $-0.5$~dB (Geisler), $-1.9$~dB (SPGD), and $1.9$~dB (DD-LMS). DD-LMS is the strongest benchmark in this test, with a mean best SNR of 5.01~dB across all trials, but it degrades beyond approximately 100~rad RMS. SPGD and Geisler's method rely on perturbation-based gradient estimates and block-averaged phase estimates, respectively. CMA/RDE optimizes an indirect modulus-error criterion. These mechanisms are less effective when large relative phase excursions disperse the pre-alignment combined power.

\subsection{Common Carrier-Frequency Offset Tolerance}
\label{sec:fo_tolerance}

Frequency-offset tolerance is important in satellite-to-ground coherent optical links~\cite{Kaushal2017}. Relative radial motion introduces an optical Doppler shift of approximately $\Delta f_{\mathrm{D}}=v_r/\lambda$, where $v_r$ is the radial velocity and $\lambda$ is the optical wavelength. At 1550~nm, radial velocities on the order of kilometers per second can correspond to carrier shifts on the order of gigahertz. Orbital prediction and coarse pre-compensation can remove most of this offset, but residual Doppler error, laser-frequency mismatch, and estimation uncertainty still require a wide acquisition range in the receiver DSP. The present simulation treats the residual offset as constant over one processing frame and does not include time-varying Doppler-rate tracking.

In a coherent receiver, residual Doppler shift and transmitter--LO frequency mismatch introduce a linear phase ramp in the complex baseband signal. Because all apertures share the same LO in the simulated receiver, the offset is common to all branches and is applied together with the LO phase noise as $\exp\!\bigl(j[2\pi \Delta f t + \phi_{\text{LO}}(t)]\bigr)$. This common phase rotation does not change the relative phases among apertures, so it does not change the BGAPA alignment objective. It must still be removed from the combined signal before symbol detection.

In this test, frequency-offset estimation (FOE) is performed at 2~samples per symbol (sps) before matched filtering. For the fourth-power estimator, the unambiguous range is $\lvert\Delta f\rvert < F_s/8$~\cite{Savory2010}; with $F_s=2R_s$ and $R_s=25$~Gbaud, this gives $\lvert\Delta f\rvert <6.25$~GHz, twice the range obtained at baud-rate sampling. BGAPA first combines the unfiltered aperture signals at 2~sps. FOE then compensates the common offset, after which the matched filter suppresses out-of-band noise before downsampling and Viterbi--Viterbi carrier phase recovery.

The common carrier-frequency offset is swept from $-12$ to $12$~GHz in 1-GHz increments at $N_{\text{ap}} = 16$, $A_{\text{phase}} = 5$~rad, $f_{\text{max}} = 1$~MHz, and $\mathrm{OSNR} = 1$~dB. For this test, both the transmitter and LO laser linewidths are set to 100~kHz. Each operating point contains $L=2^{14}$ QPSK symbols and is averaged over 20 independent Monte Carlo trials, while the BGAPA step size is individually calibrated for each $\Delta f$ value.

\begin{figure}[!htbp]
\centering
\includegraphics[width=\linewidth]{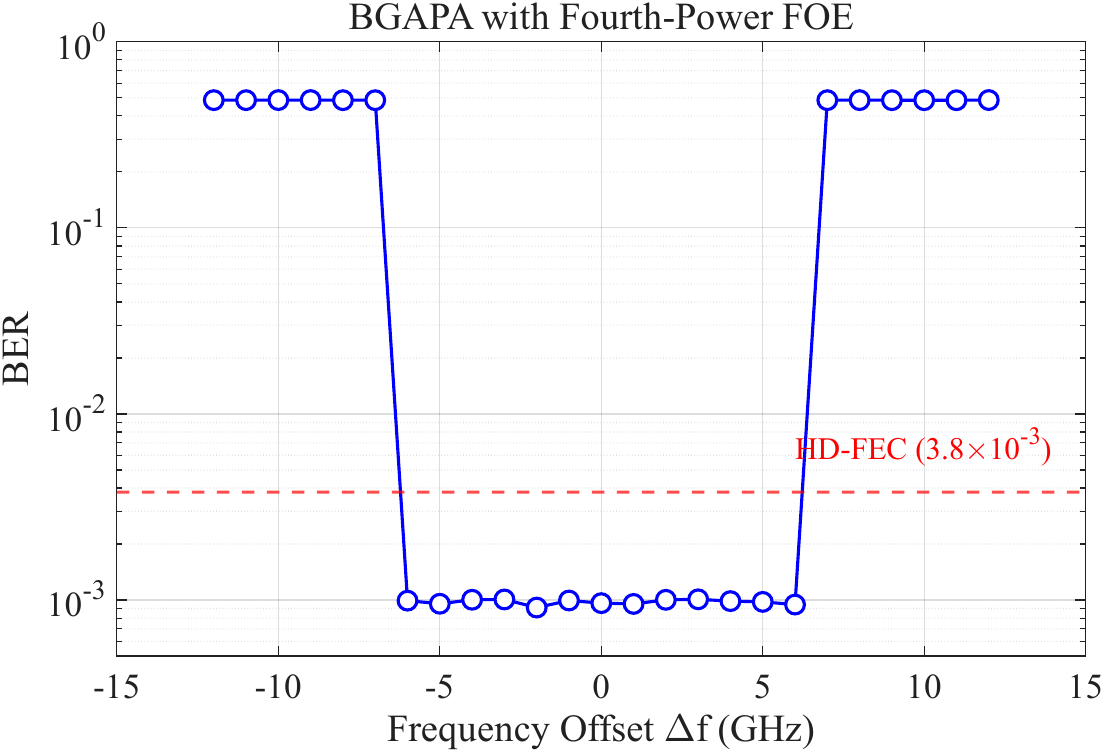}
\caption{Mean BER versus common carrier-frequency offset $\Delta f$ for BGAPA with 2-sps fourth-power FOE and matched filtering after compensation. $N_{\text{ap}} = 16$, $A_{\text{phase}} = 5$~rad, $f_{\text{max}} = 1$~MHz, $\mathrm{OSNR} = 1$~dB, and 20 Monte Carlo trials per offset. The dashed red line marks the HD-FEC threshold at $3.8\times10^{-3}$.}
\label{fig:foffset_ber}
\end{figure}

Fig.~\ref{fig:foffset_ber} shows the BER after BGAPA combining and 2-sps FOE. For $\lvert\Delta f\rvert\leq 6$~GHz, the mean BER lies between approximately $9.1\times10^{-4}$ and $1.01\times10^{-3}$, below the HD-FEC threshold. For $\lvert\Delta f\rvert\geq 7$~GHz, the BER rises to approximately 0.486 because the offset exceeds the theoretical $\pm6.25$-GHz unambiguous range of the fourth-power estimator. The observed boundary is set by the FOE capture range rather than by the relative-phase alignment objective.

\section{Conclusion}
\label{sec:conclusion}

We have proposed and numerically evaluated BGAPA for multi-aperture coherent digital combining under synthetic aperture-dependent phase disturbance. The method maximizes the combined output power through closed-form phase gradients and updates one phase parameter per aperture without training symbols, pilots, or decision-directed feedback. Across the tested aperture counts, BGAPA gives the highest post-CPR SNR among the five evaluated schemes. When the aperture count is increased by a factor of four, its SNR improvement is closest to the ideal 6.02~dB coherent-combining gain. The OSNR sweep gives the lowest HD-FEC crossing OSNR under the tested conditions. With 2-sps fourth-power FOE and matched filtering after compensation, the receiver remains below the HD-FEC threshold for $\lvert\Delta f\rvert\leq6$~GHz. Randomized disturbance-frequency and disturbance-amplitude tests further show stable BGAPA behavior in the controlled phase-only model, including a standard deviation of 0.14~dB at $N=64$ across sampled $f_{\max}$ values from 1.4 to 98~MHz and an almost flat post-CPR SNR across randomly sampled $A_{\text{phase}} \in [2,500]$~rad at $N=16$ and $f_{\max}=1$~MHz. These findings support the use of BGAPA for phase alignment in phase-dominant strong-disturbance conditions, within the limits of the present model. They are obtained with per-operating-point parameter optimization and do not establish performance under a complete atmospheric channel containing scintillation, polarization effects, and physically derived turbulence statistics. Future work will evaluate fixed-step operation, physical turbulence channels, tracking of dynamically varying polarization-state rotations, and extensions to joint phase and polarization alignment.

\bibliographystyle{IEEEtran}
\bibliography{IEEEabrv,library}

\end{document}